\documentclass[aps,prb,twocolumn,superscriptaddress,showpacs]{revtex4}
\usepackage{graphicx}
\usepackage{latexsym}
\usepackage{amssymb}
\usepackage{amsmath}
\usepackage{amsfonts}
\usepackage{bm}
\usepackage{multirow}
\usepackage{color}
\usepackage{comment}

\newcommand{\beq}{\begin{equation}}
\newcommand{\eeq}{\end{equation}}
\newcommand{\beqn}{\begin{eqnarray}}
\newcommand{\eeqn}{\end{eqnarray}}
\newcommand{\nn}{\nonumber}
 \renewcommand{\d}{\partial}

\DeclareMathAlphabet{\mathbbold}{U}{bbold}{m}{n}

\begin{document}

\title{Duality and bosonization of $(2+1)d$ Majorana fermions}

\author{Max A. Metlitski}

\affiliation{Perimeter Institute for Theoretical Physics,
Waterloo, ON N2L 2Y5, Canada}
\affiliation{Kavli Institute for Theoretical Physics, Santa Barbara, CA 93106, USA}

\author{Ashvin Vishwanath}

\affiliation{Department of Physics, Harvard University, Cambridge,
MA 02138, USA}

\author{Cenke Xu}

\affiliation{Department of Physics, University of California,
Santa Barbara, CA 93106, USA}

\begin{abstract}

We construct a dual bosonized description of a massless Majorana
fermion in $(2+1)d$. In contrast to Dirac fermions, for which a
bosonized description can be constructed using a flux attachment
procedure, neutral Majorana fermions call for a different
approach. We argue that the dual theory is an $SO(N)_1$
Chern-Simons gauge theory with a critical $SO(N)$ vector bosonic
matter field ($N\geq 3$). The monopole of the $SO(N)$ gauge field
is identified with the Majorana fermion. We provide evidence for
the duality by establishing the correspondence of adjacent gapped
phases and by a parton construction. We also propose a
generalization of the duality to $N_f$ flavors of Majorana
fermions, and discuss possible resolutions of a caveat associated
with an emergent global $Z_2$ symmetry. Finally, we conjecture a
dual description of an $\mathcal{N} = 1$ supersymmetric fixed
point in $(2+1)d$, which is realized by tuning a single flavor of
Majorana fermions to an interacting  (Gross-Neveu)  critical
point.

%to the Gross-Neveu-Yukawa model of a single flavor Majorana fermion coupled to a scalar.

\end{abstract}

\pacs{}

\maketitle

\section{Introduction}

Recently, there has been a resurgence of interest in dualities in
$(2+1)d$ quantum field theories
\cite{son2015,wangsenthil1,maxashvin,mross,karchtong,seiberg1},
building on prior work on particle-vortex dualities of bosons
\cite{peskindual,halperindual} and dualities between Dirac
fermions and bosonic theories
\cite{PolyakovAM,wufisher,barkeshlimcgreevy,minwalla,aharony1,aharony2,aharony3}.
In this paper we will discuss how analogous dualities for Majorana
fermions can be constructed.

Let us first recall the well known $(1+1)d$ case of the quantum
Ising model which admits a dual description in terms of fermionic
variables. The fermions may be viewed as the bound state of a
local spin flip and a domain wall topological defect. These new
variables allow for a direct solution of the Ising model in terms
of free Majorana fermions. In particular, the critical point
separating the Ising ordered and disordered phases can be
described either as a gapless Majorana fermion, or in terms of
bosonic variables as a critical point of a real scalar field.
Schematically:
$$
(\partial_\mu \phi)^2 + \phi^4 = {\mathcal L}^{(1+1)d}_{b} \,
\leftrightarrow \, {\mathcal L}^{(1+1)d}_f = \bar{\xi}\gamma_\mu
\partial_\mu \xi
$$
which is understood to hold in the infrared (long distance/time)
limit and the absence of mass terms on both sides represent tuning
to the critical point. This may be viewed as a bosonization of
$(1+1)d$ free Majorana fermions. How does this generalize to 2+1
dimensions? In $(2+1)d$, a Majorana mode $ {\mathcal L}^{(2+1)d}_f
$ when described in terms of a bosonic field will require a
Chern-Simons (CS) gauge field to implement the statistical transmutation. We
will argue that an $SO(N)$ gauge theory coupled to a bosonic matter
field, and supplemented by a level one CS term provides
the requisite dual description. In the dual language, the
monopoles of the gauge theory, which have a $Z_2$ character, will
be identified with the original Majorana fermions.

\subsection{Review of bosonic dual of $(2+1)d$ Dirac fermions}

It is well known that in $(2+1)d$ a CS gauge theory allows a ``bosonic" description of fermions. %For $2+1d$ Dirac fermions, a dual ``bosonic" description has been long  known.
%theory and free fermions can be generalized to $(2+1)d$ Dirac
%fermions.
A CS gauge theory binds together a
particle and a gauge flux, thereby changing the statistics of the
particle. This flux-attachment procedure certainly applies when
the particles are gapped, but it may also imply a dual description
of certain critical points. For instance, it was proposed that a
single two-component Dirac fermion in $(2+1)d$ \beq {\cal L}_\psi
= \bar{\psi} \gamma_\mu (\partial_\mu - i A_\mu) \psi + m
\bar{\psi}\psi  \label{Lpsi}\eeq has an equivalent description
\cite{PolyakovAM,wufisher,barkeshlimcgreevy,karchtong,seiberg1}:

\beqn  {\cal L}_\phi &=& |(\partial_\mu - i a_\mu) \phi|^2 + r
|\phi|^2 + g |\phi|^4 \nonumber \\
&+& \frac{i}{4\pi} (a - A) d (a - A) + 2 {\rm CS}_g\label{Lphi}
\eeqn where $\phi$ is  a complex boson, $a_\mu$ is a dynamical
$U(1)$ gauge field, and $A_\mu$ is a background $U(1)$ gauge field.
The term ${\rm CS}_g$ (with coefficient $1$) is a gravitational
Chern-Simons term (see appendix \ref{sec:CS}) encoding chiral central charge
$c_- = 1/2$. The Dirac fermion $\psi$ in ${\cal L}_\psi$ is
regularized in such a way that for $m > 0$ one gets a trivial
insulator with $\sigma^A_{xy} =  c_- = 0$, and for $m < 0$ a Chern
insulator with Chern-number 1, $i.e.$ $\sigma^A_{xy} = c_- = 1$.
Such a regularization can be obtained by starting with two gapless
Dirac cones on a $2d$ lattice and gapping one of them out. %The half-level Chern-Simons term of
%the background field $A$ in the second equation above, though does not
%affect the dynamics of the theory, comes from another massive
%Dirac fermion that must exist if this theory is regularized on a
%$2d$ lattice.

The strong interpretation of the duality is that ${\cal L}_\phi$
in Eq.~(\ref{Lphi}) at $r = 0$ flows in the infra-red to a
conformal field theory which we call the
$U(1)$-Wilson-Fisher-Chern-Simons (WFCS) fixed point, and it is
identical to a single noninteracting massless Dirac fermion ${\cal
L}_\psi$ at $m = 0$.  Though not proven, this conjecture is
consistent with various observations:

$(i)$ The Hilbert spaces and symmetries of the two theories match.
The operator dictionary between the theories ${\cal L}_\phi$ and
${\cal L}_\psi$ is as follows: the electric current of fermions
$\bar{\psi} \gamma^{\mu} \psi$ maps to the flux current of $a$,
$\frac{1}{2\pi} \epsilon^{\mu \nu \lambda} \d_{\nu} a_\lambda$.
Moreover, the fermion operator $\psi$ is the flux $2\pi$ space-time monopole
of $a$, which indeed creates $A$ charge of $1$. To see that this
object is a fermion, imagine that the system is placed on a sphere
$S^2$, with a uniform flux $2\pi$ of $a$ piercing the sphere.
Because of the self-CS term for $a$, this configuration carries
$a$ charge $1$ (in addition to  $A$ charge $-1$). Since $a$ is a
dynamical gauge field, any state in a finite volume must carry
zero $a$ charge. The $a$ charge can be neutralized by adding a
boson $\phi$ to the system. Because of the presence of the flux,
the angular momentum of $\phi$ will be half-odd-integer. Thus,
this state has the same $A$ charge and angular moment as the fermion
$\psi$.

$(ii)$ The mean field phase diagram of (\ref{Lphi}) matches the
phase diagram of the Dirac fermion (\ref{Lpsi}). Let's recall the
phase diagram of ${\cal L}_\phi$. For $r > 0$, $\phi$ is gapped,
and $a$ is gapped due to the CS term. Since the level of CS term
of $a$ is $1$, there is no intrinsic topological order: $a$
attaches flux $2\pi$ to $\phi$ turning it into a fermion, which is
identified with $\psi$ - so there are no anyons present.
Furthermore, integrating $a$ out, we obtain the Hall conductivity
$\sigma^A_{xy} = 0$, and also see that the background
gravitational CS term in ${\cal L}_\phi$ is precisely cancelled
out, so the final gravitational response is $c_- = 0$. Thus, the
$r > 0$ phase exactly matches the trivial insulator realized by
${\cal L}_\psi$ for $m > 0$. Turning to $r < 0$, the $\phi$ field
condenses, so a Higgs mass will be generated for $a$; at low
energy, one may effectively set $a_{\mu} = 0$ in ${\cal L}_\phi$.
Thus, we are left with a level $k = 1$ CS term  for the background
gauge field $A$ and a gravitational response with $c_- = 1$.  This
is a real Chern insulator, which matches the $m < 0$ phase of
${\cal L}_\psi$.

\subsection{Parton approach to Dirac duality} Let us take a $2d$
lattice model of spinless fermion $c_j$ which carries a $U(1)$
global symmetry that we label $U(1)_A$, and represent the fermion
$c_j$ with the standard parton (slave particle) construction: $c_j
= f_j b_j$, where $f_j$ and $b_j$ are slave fermion and boson
operators. All operators have the correct commutation relation,
with the following constraint on the Hilbert space on every site:
$f^\dagger_j f_j + b^\dagger_j b_j = 1$. Besides the $U(1)_A$
global symmetry, $f_j$ and $b_j$ must also carry a $U(1)$ gauge
symmetry $U(1)_a$ due to the gauge constraint: \beqn U(1)_A &:&
c_j \rightarrow e^{i\theta} c_j, \ \ \ f_j \rightarrow e^{i\theta}
f_j, \cr\cr U(1)_a &:& b_j \rightarrow e^{i\theta'} b_j, \ \ \ f_j
\rightarrow e^{- i\theta'} f_j. \label{U1transf}\eeqn Now we
design the mean field band structure of $f_j$ to be a Chern
insulator with Chern number $1$. Let us assume that the mean field
band structure of $f_j$ is always gapped, so we can safely
integrate out $f_j$, and generate a CS term for $a - A$ at level
1, as well as a gravitational CS term corresponding to $c_- = 1$,
as in ${\cal L}_\phi$ in Eq.~(\ref{Lphi}). The slave boson $b_j$
can be either in a Mott insulator, or condense. Coarse-graining
$b_j$ into a continuum field $\phi$, the transition between these
two phases is described by ${\cal L}_\phi$. While we have already
discussed the bulk phase diagram of ${\cal L}_\phi$, the parton
construction gives us additional insight into edge physics. At the
mean field level, a Chern insulator of $f_j$ has a chiral edge
mode with chiral central charge $c_- = 1$. When $b_j$ is gapped,
after coupling the edge to the dynamical $U(1)$ gauge field $a$,
there are no gauge invariant degrees of freedom left at the
boundary (or in technical terms the coset conformal field theory
at the boundary is trivial): this is consistent with the
conclusion that the $r > 0$ phase is a trivial insulator of the
physical fermion $c_j$. On the other hand, when $b_j$ condenses,
the gauge field $a$ is Higgsed, so the edge mode of $f_j$
survives; in fact, when $b$ is condensed,  $c_j \sim f_j$, so this
is just a $c_- = 1$ mode of $c_j$ as should be present in a
Chern-insulator.

The dual description (\ref{Lphi}) of a free Dirac fermion can be
generalized\cite{aharony2,seiberg2} to a $U(N)$ gauge theory,
where $a$ is a $U(N)$ CS gauge field at level $1$ and $\phi$ is a
complex boson transforming in the fundamental of $U(N)$.
Surprisingly, this theory is conjectured to be dual to a single
free Dirac cone (\ref{Lpsi}) for {\it any} $N$.\cite{seiberg2} We
review the details of this duality in appendix \ref{sec:UN}. In fact, the
dual description of a {\it Majorana} cone that we conjecture in
this paper very closely parallels this $U(N)$ dual description of
a Dirac cone. Note that this particular $U(N)$ duality is one
instance of dualities between CS-matter theories with nonabelian
gauge field and general level $k$, the duality exchanges the rank
and level of the gauge group, as well as fermionic and bosonic
matter. In the large-$N$ limit with fixed $N/k$ explicit
analytical calculations at the critical point give very strong
support for such dualities.\cite{aharony2,aharony3}

\section{Duality of a single Majorana fermion}
The main purpose of this paper is to propose a bosonized dual
description of a single two-component massless
Majorana fermion in $(2+1)d$, %Our arguments for the dual bosonized description will exactly
\begin{equation}
{\mathcal L}_\xi =\bar{\xi}\gamma_\mu \partial_\mu \xi  + m \bar{\xi}\xi. \label{Lchi}
\end{equation}
Let us consider the two phases of the Majorana fermion on tuning
the mass term $m$ from positive to negative value. We regularize
(\ref{Lchi}) so that for $m > 0$, $\xi$ realizes a trivial phase
with $c_-=0$, while for $m < 0$ it realizes a $p_x+ip_y$
superconductor with $c_- = 1/2$. Such a regularization can be
provided by starting with a lattice model with two gapless
Majorana cones and initially gapping one of them out to produce
(\ref{Lchi}) with $m  =0$.  Thus, the massless Majorana cone
corresponds to the critical point between a trivial phase and a
$p_x+ip_y$ superconductor. Now, the idea is to produce a dual
theory that realizes the same two phases on tuning a parameter.
Then at the critical point we may conjecture the dual theory has
the same infrared behavior as (\ref{Lchi}), given the paucity of
available fixed points. Observe that we can also capture the same
pair of gapped phases with the following theory of a boson coupled
to an $SO(N)_1$ CS gauge field:
%We include the response of a gravitational Chern Simons field at level N (corresponding to `N' chiral Majorana modes) in the effective theory:
\begin{equation}
{\mathcal L}_b = |(\partial_\mu
- i a_\mu) \phi|^2 + r |\phi|^2 + g |\phi|^4 + \mathrm{CS}_{SO(N)}[a]_1+N \cdot \mathrm{CS}_g \label{SON}
\end{equation}
Here, $\phi$ is an $N$-component real vector, $a$ is an $SO(N)$
gauge field and \beq {\rm CS}_{SO(N)}[a]_1 = \frac{i}{2\cdot 4\pi}
\mathrm{tr}_{SO(N)}\left(a\wedge da - \frac{2 i}{3} a \wedge a
\wedge a\right). \label{CSO}\eeq The trace in Eq.~(\ref{CSO}) is in
the vector representation of $SO(N)$ (for a more precise
definition of the CS term, see appendix \ref{sec:CS}).

Let's analyze the mean field phase diagram of (\ref{SON}). When
$r>0$, $\phi$ is gapped. The theory $SO(N)_1$ coupled to gapped
bosonic matter gives a state with no intrinsic topological order;
the vector boson $\phi$ is transmuted to a fermion by the CS
field.\footnote{The $SO(3)_1$ example might be the most familiar:
this theory is the same as $SU(2)_2 = \{1, \sigma, f\}$ restricted
to integer spin, $i.e.$ to $\{1, f\}$. The chiral central charge is
$c_-= -3/2$.}  The $SO(N)_1$ CS gauge theory by itself has a
chiral central charge $c_- = -N/2$, which exactly cancels the
background gravitational CS term in Eq.~(\ref{SON}). So the $r >
0$ phase is a trivial state with $c_- = 0$. On the other hand, for
$r < 0$, $\phi$ condenses, which breaks the gauge group $SO(N)$ to
$SO(N-1)$. At low energies, we can then take $a$ to be an
$SO(N-1)$ gauge field, obtaining an $SO(N-1)_1$ CS theory. Again,
this is a state with no intrinsic topological order, but the
background gravitational term in (\ref{SON}) is no longer fully
cancelled, rather: $c_- = N/2 - (N-1)/2 = 1/2$. So the $r < 0$
phase is a $p_x+ip_y$ superconductor.

\subsection{Parton approach to Majorana duality}

To further motivate the dual theory (\ref{SON}) we utilize a
parton construction. The simplest construction is: $c =
\sum_{\alpha = 1}^{N} \tilde{\phi}_\alpha \tilde{\chi}_\alpha$,
where the Majorana parton $\tilde{\chi}_\alpha$ and slave boson
$\tilde{\phi}_\alpha$ have an $O(N)$ redundancy. However, to
recover the $SO(N)$ gauge structure from this construction
requires further discussion, which we provide in appendix \ref{app:Z2}.
Instead, here we turn to a slightly different parton
representation. Consider a lattice model for Majorana fermion
$c_j$ ($c_j^\dagger = c_j$). This time we introduce  on
each site $j$, $N$ colors of
slave Majorana fermions $\chi_{j,\alpha}$ ($\alpha = 1 \ldots N$) for odd  $N$ such that
\beqn c_j =
(i)^{\frac{N-1}{2}} \prod_{\alpha = 1}^N \chi_{j,\alpha}.
\label{parton1}\eeqn By construction $\chi_{\alpha}$ is coupled to
a dynamical $SO(N)$ gauge field. Now we design an identical mean
field $p_x+ip_y$ superconductor band structure for each color of
$\chi_\alpha$. At the mean field level, there are $N$ chiral
Majorana fermions at the boundary, which in total leads to chiral
central charge $c_- = N/2$. However, if the $SO(N)$ gauge symmetry
is unbroken, after coupling to the $SO(N)$ gauge field there will
be no gauge invariant degrees of freedom left at the boundary, so
we are left with $c_- =0$. Also, integrating out $\chi_{\alpha}$
would generate a CS term at level $k = 1$ for the $SO(N)$ gauge
field, as well as a gravitational CS term at level $N$, as in
Eq.~(\ref{SON}). As already discussed, there is no topological
order in the bulk, thus this state is again a trivial state of the
physical Majorana fermion $c_j$.

Using the slave particles $\chi_{j,\alpha}$ we can also define a
$SO(N)$ vector boson $ \hat{\phi}_{j,\alpha}$ \footnote{Strictly
speaking, $\hat{\phi}_\alpha$ satisfy $\prod_{i = 1}^{N}
\hat{\phi}_\alpha = 1$, but this constraint is assumed to be
softened in the infrared.}: \beqn \hat{\phi}_{j, \alpha} \sim
(i)^{\frac{N-1}{2}} \epsilon_{\alpha \alpha_1, \cdots
\alpha_{N-1}} \chi_{j, \alpha_1} \cdots \chi_{j,\alpha_{N-1}}.
\eeqn
%$\hat{\phi}_{j,\alpha}$ obeys the %SO($N$) Clifford algebra.
When $\hat{\phi}_{\alpha}$ condenses, it breaks the $SO(N)$ gauge
group down to $SO(N-1)$, and one of the slave fermions (say
$\chi_N$) is no longer coupled to any gauge field, thus its
topological band structure implies that the entire system is
equivalent to one copy of $p_x + ip_y$ topological superconductor.
Likewise, the edge mode associated to $\chi_N$ sees no gauge field
and survives as a true $c_- = 1/2$ edge mode of a $p_x + i p_y$
superconductor. All other edge modes are, as before, eliminated by
$SO(N-1)$ gauge field fluctuations. Thus, by coarse-graining
$\hat{\phi}$ into a continuum field $\phi$, we can describe a
transition between a trivial state and a $p_x + ip_y$
superconductor by Eq.~(\ref{SON}).

Notice that the integer $N$ in the dual theory (\ref{SON}) needs
not be odd. We could adjust our mean field construction, and take
the band structure of $\chi_1, \cdots \chi_{N-1}$ to be an
identical $p_x + i p_y$ superconductor, while placing $\chi_N$
into a trivial band structure. Then the mean field band structure
already breaks the $SO(N)$ gauge symmetry to $O(N-1)$ gauge
symmetry, and $\hat{\phi}_{\alpha}$, $\alpha = 1\ldots N-1$,
reduces to a $SO(N-1)$ vector. The extra $Z_2$ subgroup of the
$O(N-1)$ gauge symmetry can be broken by condensing the $SO(N-1)$
gauge singlet bosonic operator $\hat{\phi}_N =
(i)^{\frac{N-1}{2}}\prod_{\alpha = 1}^{N-1} \chi_\alpha$. Now
condensing the $SO(N-1)$ vector $\hat{\phi}_{\alpha}$ also drives
a transition from the trivial state of $c_j$ to a $p_x + ip_y$
superconductor of $c_j$, and this time the transition is described
by a $SO(N-1)$ vector field coupled to a $SO(N-1)$ gauge field
with a CS term at level $1$.

% and we again propose that the infrared physics of
%this theory is a CFT that is dual to a single Majorana fermion,
%which describes the same transition.

%This phase transition described in terms of the
%corresponding coarse-grained field $\phi_{\alpha}$ of the operator
%$\hat{\phi}_{\alpha}$ is \beqn |(\partial_\mu - i a_\mu) \phi|^2 +
%r |\phi|^2 + g |\phi|^4 + \mathrm{CS}[a]_1, \label{onwfcs}\eeqn
%where $a_\mu$ is the dynamical SO($N$) gauge field, and
%$\mathrm{CS}[a]_1$ is a Chern-Simons term at level $ k = 1$ for
%$a_\mu$.

%As we argued above, this transition is essentially a transition
%from a trivial superconductor to a topological superconductor, thus it
%can also be described by a single two-component Majorana fermion.
%This observation prompts us to propose the following dual
%description of a single Majorana fermion: \beqn && |(\partial_\mu
%- i a_\mu) \phi|^2 + r |\phi|^2 + g |\phi|^4 + \mathrm{CS}[a]_1
%\cr\cr &\leftrightarrow& \bar{\chi}\gamma_\mu \partial_\mu \chi +
%m \bar{\chi}\chi. \label{dual2}\eeqn Notice that there must be
%another massive Majorana fermion in the system, as long as the
%system is regularized on a $2d$ lattice. {\color{red} some rewriting here that needs to be fixed according to the new flow} Our prediction is that,
%the infrared physics of the first line of Eq.~\ref{dual2} at $r =
%0$ is identical to the second line at $m=0$. For instance, at the
%critical point $r = 0$, the duality dictates that the scaling
%dimension of $|\phi|^2$ is 2.

\subsection{The dictionary}
How do we represent the physical Majorana fermion in the dual
theory (\ref{SON})? In the case of Dirac duality (\ref{Lpsi})
$\leftrightarrow$ (\ref{Lphi}) the physical electric charge in the
Dirac theory mapped to the magnetic flux of the $U(1)$ gauge field
$a$ in the dual theory. Likewise, the fermion parity of the
Majorana fermion $\xi$ maps to the magnetic flux of the $SO(N)$
gauge field $a$ in Eq.~(\ref{SON}). Recall that the magnetic flux
is classified by $\pi_1(SO(N)) = Z_2$. Indeed, imagine the system
on a spatial sphere $S^2$. As usual, we place a magnetic flux
through the sphere by dividing it into two hemispheres, and gluing
the fields in the two hemispheres along the equator with a gauge
transformation $g(\theta)$, $\theta \in [0, 2 \pi]$. Such gauge
transformations are classified by $\pi_1$ of the gauge group. In
the case of the $SO(N)$ group, a simple representative for the
single non-trivial magnetic flux sector on $S^2$ is obtained by
considering an ordinary flux $2 \pi m$ Dirac monopole in the
$SO(2)$ subgroup of $SO(N)$ with $m  =1$. Note that by an $SO(N)$
rotation we can invert the magnetic flux in the $SO(2)$ subgroup,
so $m$ is, indeed, only defined modulo $2$. The magnetic flux
breaks the $SO(N)$ group down to $SO(2) \times SO(N-2)$ and the
state on $S^2$ must be neutral under this reduced gauge  group. As
in the abelian case, the CS term in Eq.~(\ref{SON}) leads to the
monopole carrying an $SO(2)$ charge $1$, so to make the monopole
neutral we must act on it with the boson $\phi_1 + i \phi_2$. As
before, the boson angular momentum is  half-odd-integer  because
of the $SO(2)$ flux, so the angular momentum of the resulting
state is half-odd-integer. We conclude that the $SO(N)$ monopole
on $S^2$ carries charge under fermion parity, and identify the
$SO(N)$ space-time monopole $V_M$ with the Majorana fermion operator $\xi$.
In particular, this discussion means that dynamical $Z_2$ $SO(N)$
monopoles are prohibited in the partition function of dual
theory (\ref{SON}), as they violate fermion parity conservation.

One can also see that the $SO(N)$ monopole on $S^2$ will have a
non-trivial fermion parity from the parton construction. Indeed,
when there is a $2\pi$ flux of $SO(2)$ through the sphere, it will
be seen by  partons $\chi_1, \chi_2$, while $\chi_\alpha$, $\alpha
= 3\ldots N$ will see no flux. The ground state will then have
$SO(2)$ charge $1$. Since $\chi_\alpha$ carry fermion parity, the
ground state also carries $(-1)^F = -1$. The $SO(2)$ charge gets
neutralized by adding a boson $\phi_1 + i \phi_2$. However, since
$\phi$'s carry no fermion parity, $(-1)^F = -1$ is not affected.

More generally, for a spatial manifold $\Sigma$, a gauge field
configuration $a$ and boundary conditions (spin-structure)
$\sigma$ for the physical Majorana fermion $\xi$, the fermion
parity in the dual theory (\ref{SON}) is expressed
as\cite{Jenquin1, Jenquin2}: \beq (-1)^F = (-1)^{\int_\Sigma W_2}
\, (Arf[\Sigma, \sigma])^N \label{W2}\eeq Here, $W_2 \in
H^2(\Sigma, Z_2)$ is the obstruction to lifting the $SO(N)$ gauge
bundle to $Spin(N)$. $Arf(\Sigma, \sigma) = \pm 1$ is the $Arf$
invariant, which computes the fermion parity of a $p_x+ip_y$
superconductor on $\Sigma$ with spin-structure $\sigma$.\cite{KapustinF} Thus,
only the first term in (\ref{W2}) depends on the gauge field, the
second term is fixed once the boundary conditions for $\xi$ are
fixed. The first and second terms in Eq.~(\ref{W2}) are
correspondingly secretly encoded in the spin-structure dependence
of CS action (\ref{CSO}) and the gravitational CS term in
(\ref{SON}).
% The second term on the RHS of (\ref{W2}) is independent of the $SO(N)$ gauge field and is sensitive ,

Having established the equivalence of phases and operators in the
free Majorana theory (\ref{Lchi}) and the dual theory (\ref{SON}),
we conjecture that they are actually dynamically equivalent at
their respective IR fixed points.
%For a SO($N$) gauge group with $N
%\geq 3$ we note that $\pi_1\left [ \mathrm{SO}(N) \right ]
%=Z_2$. %This is relevant to classifying the monopoles on the gauge
%theory (since these correspond to the allowed generalizations of
%Dirac strings \cite{Preskill}).
%Let the monopole creation operator (equivalent to the destruction
%operator) be $V_M$. Due to the presence of the CS term, the
%monopole acquires SO($N$) gauge charge. In order to neutralize
%this, let us bind it to a $\phi_\alpha$ boson. The resulting
%object $\chi = \phi \cdot V_M$ is gauge neutral and is a fermion
%(Since it is the bound state of unit monopole and gauge charge).
%Because for $N \geq 3$ the monopole of SO($N$) gauge field is the
%same as its anti-monopole, we identify $\chi$ as the Majorana
%fermion operator.

\begin{table}[htp]
\caption{Duality Dictionary}
\begin{center}
\begin{tabular}{|c|c|}
\hline
Fermionic Theory & Bosonic Theory \\
\hline

Majorana: $\xi$ & Monopole: $V_M$ \\
``m": $\bar{\xi}\xi$ & ``r": $\phi \cdot \phi $\\
\hline
\end{tabular}
\end{center}
\label{default}
\end{table}%
We note that the level-rank duality of Chern-Simons-matter
gauge theories with $O(N)_k$ gauge group in the large-$N$, large-$k$ limit has been proven in
Ref.~\onlinecite{aharony2}. Our conjecture (\ref{Lchi}) $\leftrightarrow$ (\ref{SON}) amounts to a statement that the duality continues to hold when $k  =1$ and $N$ is finite (with a clarification  that the precise form of the gauge group for $k  =1$ is $SO(N)$).  %Based on the arguments
%presented above, we believe this duality is also valid for
%$SO(N)_1$ Chern-Simons matter field theory.

%{\color{red} Can we say something about the 3+1D theory that leads to this duality? For example, lets specialize to $N=3$ for a second. Begin with a 3D SPT of SO(3) symmetric bosons with time reversal symmetry. Then, there is a boson SPT such that the weakly gauged SO(3) monopole is a Majorana fermion.

%}
\subsection{$Z_2$ global symmetry of the bosonic theory}
One potential subtlety of the proposed duality (\ref{Lchi})
$\leftrightarrow$ (\ref{SON}) is that the dual theory actually has
a global $Z_2$ symmetry. Indeed, (\ref{SON}) is invariant under
$\phi \to  O \phi$, with $O \in O(N)$; while for $O \in SO(N)$
this is a gauge transformation, $Z_2 = O(N)/SO(N)$ remains as a
global symmetry. We can write down an ``Ising order parameter"
$\Phi$ for this global $Z_2$ symmetry,
\beqn \Phi(x) &=& \epsilon_{\alpha_1 \alpha_2 \ldots \alpha_N} (W(x, x_1) \phi(x_1))_{\alpha_1} (W(x, x_2) \phi(x_2))_{\alpha_2} \nonumber \\
&& \ldots(W(x, x_N) \phi(x_N))_{\alpha_N} \label{Phi}\eeqn where
$W(x, y)$ is an $SO(N)$ Wilson line and $\Phi(x)$ is obtained by
taking the limit $x_i \to x$ and keeping the leading surviving
terms. For instance, for $N = 3$, we have a Lorentz scalar, $\Phi =
\epsilon^{\alpha\beta\gamma} \phi_\alpha (D^2 \phi)_\beta (D^4
\phi)_\gamma$, where $D_{\mu} = \d_{\mu} - i a_{\mu}$.
%Given the number of fields and derivatives in $\Phi$ one may expect it to be an irrelevant perturbation to (\ref{SON}), so that the dual theory has an emergent $Z_2$ symmetry.
The presence of an extra global $Z_2$ symmetry in the dual theory
is problematic, since the only obvious symmetry of the Majorana
cone is fermion parity $(-1)^F$. Clearly, the $Z_2$ symmetry of
the dual theory is not $(-1)^F$ since the local boson $\Phi$
transforms non-trivially under it. We conjecture the following
resolution to this puzzle: the gap to $Z_2$ charged excitations in
(\ref{SON}) remains finite across the critical point, $i.e.$ $\Phi$
actually has exponentially decaying correlation functions at $r =
0$. Here's the evidence for this conjecture:

%must be
%an SO($N$) gauge singlet, hence it is a total antisymmetric
%combination of $\phi_\alpha$ with multiple space-time derivatives.
%{\color{red} Write explicit form}
%But this order parameter $\Phi$ must be gapped in the entire phase
%diagram, because
({\it i}) Neither phase in the phase diagram has spontaneous $Z_2$
symmetry breaking, $i.e.$ $\langle \Phi \rangle = 0$. Indeed, when
$\phi$ develops an expectation value, the $O(N)$ symmetry is
broken to $O(N-1)$ and the gauge symmetry $SO(N)$ to $SO(N-1)$, so
$Z_2 = O(N)/SO(N) \to O(N-1)/SO(N-1)$ survives.

({\it ii}) While $Z_2$ remains unbroken for both $r > 0$ and $r <
0$, these two regions could potentially realize different $Z_2$
symmetry protected phases. Closing of a $Z_2$ charge gap is then
required at the transition. However, it can be shown that both $r
> 0$ and $r < 0$ realize the same (trivial) $Z_2$ symmetry protected
topological (SPT) phase. To see this, observe that our parton
construction gives no $Z_2$ carrying edge modes on either side of
the transition, which means that both phases realize a trivial
$Z_2$ SPT (see appendix \ref{app:Z2} for further details).

Given $({\it i})$ and $({\it ii})$, there is no requirement that
the $Z_2$ charge gap closes at the transition, so we make the
minimalist assumption that it remains finite and the effect of
$Z_2$ global symmetry can be ignored in Eq.~(\ref{SON}),
consistent with the proposed duality.

We note that instead of considering an $SO(N)$ gauge theory in
(\ref{SON}), we could make the gauge group $O(N)$ - $i.e.$ gauge the
extra $Z_2$ symmetry. Then the $r > 0$ phase will be a (trivial)
$Z_2$ gauge theory (toric code) and the $r < 0$ phase will be a
toric code +  a $(p_x + i p_y)$ superconductor. The assumption
that the $Z_2$ charge gap remains finite across the transition is
equivalent to the toric code being an inert spectator. The $O(N)$
gauge theory is then conjectured to be dual to a Majorana cone
(\ref{Lchi}) with a decoupled gapped toric code.

%Potentially, Z$_2$ symmetry could protect a topological phase, whose nature could change across the transition: which would mandate
%a closing of $Z_2$ charge gap at the transition. leading to gapless Z$_2$ excitations. However, this can also be shown not to occur {\color{red} See Max's note below}  ({\it iii.}) $\Phi$ cannot be a composite
%operator of the gauge invariant Majorana fermion which becomes
%gapless at the critical point, $i.e.$ $\Phi$ has no reason to be
%gapless at the critical point.

%Notice that in the duality, the integer $N$ needs not be odd. We
%could adjust our mean field construction, and take the band
%structure of $\chi_1, \cdots \chi_{N-1}$ to be an identical
%topological superconductor, while $\chi_N$ has a trivial band
%structure. Then the mean field band structure already breaks the
%SO($N$) gauge symmetry to O($N-1$) gauge symmetry, and
%$\hat{\phi}_{\alpha}$ reduces to a SO($N-1$) vector. The extra
%$Z_2$ subgroup of the O($N-1$) gauge symmetry can be broken by
%condensing the SO($N-1$) gauge singlet bosonic operator
%$(i)^{\frac{N-1}{2}}\prod_{\alpha = 1}^{N-1} \chi_\alpha$. Now
%condensing $\phi_{\alpha}$ also drives a transition from the
%trivial state of $c_j$ to a topological superconductor of $c_j$,
%and this times this transition is described by a SO($N-1$) vector
%field coupled to a SO($N-1$) gauge field with a Chern-Simons term
%at level$-1$, and we again propose that the infrared physics of
%this theory is a CFT that is dual to a single Majorana fermion,
%which describes the same transition.

\begin{figure}[tbp]
\begin{center}
\includegraphics[width=240pt]{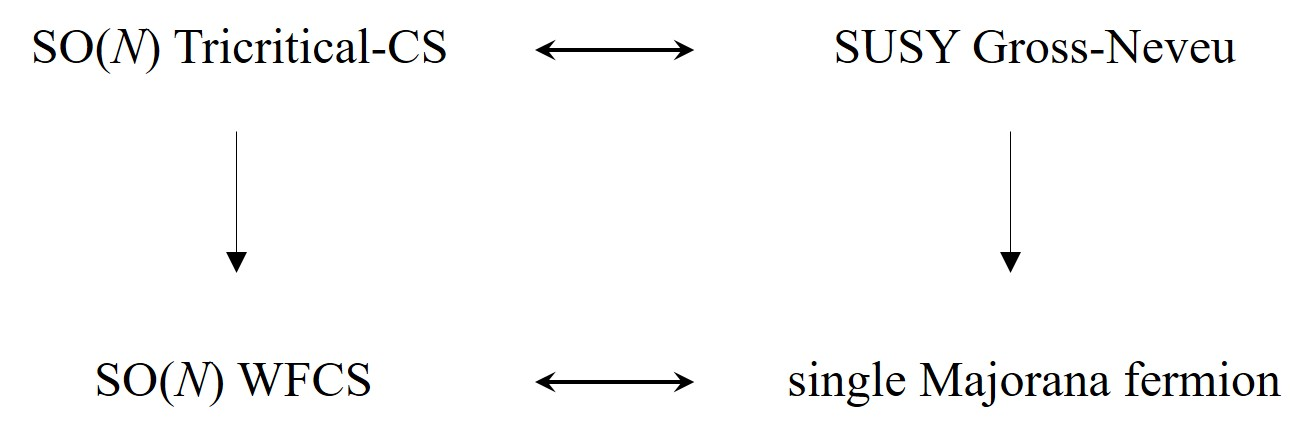}
\caption{The proposed duality and renormalization group flow. The
double headed arrow stands for ``dual to each other", while the
single headed arrow represents the RG flow.} \label{RG}
\end{center}
\end{figure}

\subsection{Dual of the Gross-Neveu-Yukawa theory}
The critical point of the $SO(N)$-WFCS theory is an infrared (IR)
fixed point of an ultraviolet (UV) fixed point, which we call the
$SO(N)$-Tricritical-CS theory. This theory corresponds to tuning
$g$ in Eq.~(\ref{SON}) to a critical value $g_c$: at mean field
$g_c = 0$  (we assume that the action is still bounded from below
due to the existence of higher order terms in the polynomial of
$|\phi|$), which is analogous to the tricritical
Ising fixed point.
% For $g > g_c$, we recover our old dual theory (\ref{SON}) where (at least at mean field) the transition as a function of $r$ is continuous, while for $g <  g_c$ the $\phi$ condensation transition is first order.

On the fermion side, a natural UV fixed point that flows to the
free Majorana fermion in the IR is the Gross-Neveu-Yukawa fixed
point: \beqn \mathcal{L} = \bar{\xi}\gamma_\mu \partial_\mu \xi +
(\partial_\mu \sigma)^2 + \lambda \sigma \bar{\xi}\xi +
\frac{\tilde{\lambda}^2}{4} \sigma^4. \label{gny}\eeqn where
$\sigma$ is a real scalar. The relevant perturbation $s \sigma^2 $
in Eq.~(\ref{gny}) is dual to $(g-g_c) |\phi|^4$ in
Eq.~(\ref{SON}): when $s, \, g-g_c > 0$, Eq.~(\ref{gny}) and the
$SO(N)$-Tricritical-CS theory respectively flow to
Eqs.~(\ref{Lchi}) and (\ref{SON}). On the other hand, when $s, \,
g-g_c < 0$, both theories have a first order transition between a
trivial phase and a $p_x + i p_y$ superconductor.

An exact renormalization flow of Eq.~(\ref{gny}) is difficult to
compute, but if there is only one fixed point with nonzero
$\lambda$ and $\tilde{\lambda}$, then this fixed point must have
$\lambda^\ast = \tilde{\lambda}^\ast$, and it is a supersymmetric
$\mathcal{N} = 1$ conformal field
theory~\cite{grovervishwanath1,grovervishwanath2,klebanov}. Thus,
our construction also conjectures that this $\mathcal{N} = 1$
supersymmetric conformal field theory is dual to the
$SO(N)$-Tricritical-CS theory. The supersymmetry makes the
following prediction about the scaling dimensions of the
$SO(N)$-Tricritical-CS theory: \beqn \Delta[V_M] -
\Delta[|\phi|^2] = 1/2.\eeqn

An analogous duality between the Dirac fermion Gross-Neveu fixed
point and the $U(1)$-Tricritical-CS fixed point was conjectured in
Ref.~\onlinecite{karchtong}.

\section{Duality of $N_f$ flavors of Majorana fermions}

There is a generalization of our proposed duality to the case of
$N_f$ Majorana fermions with $SO(N_f)$ flavor symmetry.
%Here we
%will focus on the case with an odd integer $N_f$,  because an even
%number of Majorana fermions is equivalent to $N_f/2$ Dirac
%fermions.{\color{blue} But the full $SO(N_f)$ symmetry is not
%apparent when we use the Dirac duality, so maybe better not to use
%this excuse.;)}
 %In this case, $O(N_f) = Z^F_2 \times SO(N_f)$
%where $Z^F_2$ is the fermion parity group.
Take $N_f$ Majorana fermions, \beq  \mathcal{L} = \sum_I
(\bar{\xi}_I \gamma_\mu
\partial_\mu \xi_I + m \bar{\xi}_I \xi_I )\eeq where $I = 1\ldots
N_f$. For $m > 0$, we have a trivial insulator with no $SO(N_f)$
response and $c_- = 0$; for $m < 0$, we have chiral central charge $c_- =
N_f/2$ and CS $SO(N_f)$ response at level $ k = 1$: \beq
\mathcal{L} = \mathrm{CS}_{SO(N_f)}[A]_1 + N_f \cdot \mathrm{CS}_g \label{Nfresp}\eeq where
$A$ is the background $SO(N_f)$ gauge field.% and CS$(g)$ is the
%gravitational response of a $p_x + ip_y$ superconductor ($c =
%1/2$).

We propose that the above theory is dual to $N_f$ flavors of real
bosons $\phi^\alpha_I$ coupled to an $SO(N)$ gauge field $a_\mu$
at level $ k = 1$ (as before, $\alpha = 1 \ldots N$ is the color
gauge index): \beqn \mathcal{L} &=& \sum_I |(\partial_\mu - i
a_\mu) \phi_I|^2 + r (\mathrm{tr} M) + u (\mathrm{tr}M) ^2 + v
\mathrm{tr}(M^2) \cr\cr &+& \mathrm{CS}_{SO(N)}[a]_1 + N \cdot
\mathrm{CS}_g \label{eq:Lflav} \eeqn where $M_{IJ} =
\phi^\alpha_I \phi^\alpha_J$. The physical $SO(N_f)$ flavor
symmetry simply acts on the flavor indices $I$ of $\phi^\alpha_I.$
We will assume $N \ge N_f +2$. We further assume that either $N$ is even and $N_f$ is arbitrary, or $N$ and $N_f$ are both odd.\footnote{The case of odd $N$ and even $N_f$ requires working with an $O(N)$ gauge group.} When $r > 0$, $\phi$ is not
condensed, so integrating out gauge field $a$ cancels out the
gravitational CS term and gives rise to a trivial insulator.

We assume that $v > 0$ in Eq.~\ref{eq:Lflav}.\footnote{If the
gauge field is ignored, with large $N$ and fixed $N_f$, there does
exists a stable fixed point with
$u> 0$ and $v > 0$.\cite{vicari2007}.}%{\color{blue} Is there no fixed point with $v < 0$ (was not super-clear from Vicari's paper)? At any rate, $v < 0$ suggests a different pattern of symmetry breaking, so those are really different universality classes.}
\,\, In this case, in the phase
$r < 0$, the energy is minimized when $\langle \phi^\alpha_I
\phi^\alpha_J \rangle = \langle M_{IJ} \rangle \sim \delta_{IJ}$ -
$i.e.$ this gauge invariant observable suggests that the flavor
symmetry remains unbroken (as we will see more explicitly below)
by the condensate of $\phi^\alpha_I$, or equivalently
$\langle \phi^\alpha_I \rangle$ are $N_f$ orthogonal $SO(N)$ vectors.

With $v > 0$, in the condensed phase of $\phi^\alpha_I$ we can
choose $\langle \phi^\alpha_I \rangle \sim \delta^\alpha_I$. The
gauge group $SO(N)$ is broken down to $SO(N-N_f)$ (acting on
$\alpha = N_f + 1 \ldots N$). Furthermore, the combination of an
identical $SO(N_f)$ flavor rotation and an $SO(N_f)$ color
rotation acting on $\alpha = 1\ldots N_f$ leaves $\langle
\phi^{\alpha}_I \rangle$ invariant, which means that the physical
$SO(N_f)$ global symmetry remains unbroken. In the presence of a
background $SO(N_f)$ gauge field $A_\mu$, the components of the
dynamical gauge field $a^{\alpha \beta}_\mu$ with $\alpha, \beta =
1\ldots N_f$ get Higgsed to $a^{\alpha \beta}_\mu =
A^{\alpha\beta}_\mu$, similarly $a^{\alpha \beta}_\mu = 0$ for
$\alpha = 1 \ldots N_f$, $\beta = N_f + 1 \ldots N$. Finally,
$a^{\alpha\beta}_\mu$ with $\alpha, \beta = N_f+1 \ldots N$ is not
Higgsed - let's refer to this non-Higgsed field as $b_\mu$.
Therefore, our effective action takes the form: \beq \mathcal{L} =
\mathrm{CS}_{SO(N-N_f)}[b]_1 + \mathrm{CS}_{SO(N_f)}[A]_1 + N
\cdot \mathrm{CS}_g.\eeq Integrating $b$ out, we recover
Eq.~(\ref{Nfresp}).

Just like in the $N_f = 1$ case, let us discuss a parton construction
for the case with $SO(N_f)$ flavor symmetry. On every lattice site,
we use the parton decomposition: \beq c_I =
(i)^{\frac{N}{2}}\chi_I \prod_{\alpha = 1}^N \chi^\alpha, \eeq
with even integer $N$. The site index is hidden. Under flavor
symmetry \beq O(N_f): c_I \to R_{IJ} c_J, \quad \chi_I \to R_{IJ}
\chi_J, \quad \chi^\alpha \to \chi^\alpha \label{eq:ONf} \eeq
There is also an $O(N)$ gauge symmetry: \beq O(N): \chi^\alpha \to
V^{\alpha\beta} \chi^\beta, \quad \chi^I \to ({\rm det} V) \chi^I. \eeq In order to get
Eq.~(\ref{eq:Lflav}), we still design the mean field band structure
of all $\chi^\alpha$ to be an identical $p_x + ip_y$ topological
superconductor. On the other hand, $\chi^I$ are chosen to have
identical trivial band structures. Then the phase transition of Eq.~(\ref{eq:Lflav}) is
the order-disorder transition of the bosonic operator
$\hat{\phi}^\alpha_I = i \chi_I \chi^\alpha$, which is a vector of
both $O(N_f)$ and $O(N)$.
%The definition of $\hat{\phi}^\alpha_I$
%guarantees that on every side \beqn \sum_\alpha
%\{\hat{\phi}^\alpha_I, \ \hat{\phi}^\alpha_J \} \sim \delta_{IJ}.
%\eeqn This algebra of $\hat{\phi}^\alpha_I$ suggests that its most
The field $\phi^\alpha_I$ in Eq.~(\ref{eq:Lflav}) is the
coarse-grained field of operator $\hat{\phi}^\alpha_I$.

%$N_f$ orthogonal SO($N$) vectors, which implies $v
%> 0$ in Eq.~\ref{eq:Lflav}, and hence in the condensate of
%$\phi^\alpha_I$, $\langle M_{IJ} \rangle \sim \delta_{IJ}$.

%Besides giving $\chi^\alpha$ a $p_x + ip_y$ topological
%superconductor mean field band structure,
Notice that the gauge symmetry in the above construction is
$O(N)$. To match with our definition of the continuum theory
(\ref{eq:Lflav}), we want to break it to $SO(N)$. Just like in the
$N_f = 1$, even $N$ case, we can simply condense $ \prod_\alpha
\chi^\alpha$ (and keep it condensed throughout the phase diagram
of Eq.~(\ref{eq:Lflav})). The $O(N_f)$ flavor symmetry and the
$SO(N)$ part of the gauge symmetry are preserved, but the
reflection part of the $O(N)$ gauge symmetry is broken, as needed.
A modified mean field band structure with one of $\chi^\alpha$s
forming a trivial band  allows us to realize Eq.~(\ref{eq:Lflav})
with odd $N$ (and odd $N_f$).

\section{Discussion}

Our dual theory Eq.~(\ref{SON}) obviously breaks time-reversal
symmetry, while the single Majorana cone (\ref{Lchi}) could
preserve the time-reversal symmetry, if the system is defined on
the $2d$ boundary of a $3d$ topological superconductor in class
DIII (the  topological phase with index $\nu=1$ is believed to be
realized by the B-phase of superfluid He$^3$). For the Dirac
fermion, the dual U(1)-WFCS theory (\ref{Lphi}) is believed to
have an emergent time-reversal symmetry in the IR, which
transforms the matter field $\phi$ into its vortex.\cite{seiberg1}
However, this simple solution does not apply to our $SO(N)$ gauge
theory, as there is no known analogue of the boson-vortex duality
for an $SO(N)$ matter field with $N \ge 3$. Thus, we do not yet
understand how time-reversal is hidden in the dual
theory (\ref{SON}).
%Moreover,
Moreover, in the case of the Dirac fermion, a manifestly
$T$-invariant description is provided by another dual theory:
QED$_3$ with a single Dirac fermion matter field.\cite{son2015,
wangsenthil1, maxashvin, seiberg1} Again, a manifestly
$T$-invariant dual description of a Majorana cone is currently
missing. Such a theory could provide a derivation of the surface
topological order of class DIII topological phases with odd
$\nu$, such as the proposed SO(3)$_3$ nonabelian topological order
with just two nontrivial
particles~\cite{fidkowskihe,Teo,WangLevin}. We recall that in the
Dirac context, the pairing of composite Dirac fermions led to an
explicit  derivation of the  T-Pfaffian state, a surface
topological order for the topological
insulator\cite{TI_fidkowski2, son2015,wangsenthil1,maxashvin, Ti_qi}. We
leave these questions to future work.

The authors thank Maissam Barkeshli, Meng Cheng, Chao-Ming Jian,
Mike Mulligan, Chetan Nayak and Ryan Thorngren for very helpful
discussions. Through private communication, the authors have
learned that N. Seiberg $et \,al.$ are also studying dualities of
$SO(N)$ CS-matter theory using different approaches. The authors thank the organizers of the KITP program {\it ``Symmetry, Topology, and Quantum Phases of Matter: From Tensor Networks to Physical Realizations"} during which this work was initiated.
AV was
supported by a Simons Investigator Grant. C. Xu is supported by
the David and Lucile Packard Foundation and NSF Grant No.
DMR-1151208. Research at Perimeter Institute
for Theoretical Physics (MM) is supported by the
Government of Canada through the Department of Innovation,
Science and Economic Development and by the
Province of Ontario through the Ministry of Research
and Innovation. This research was supported in part by the National Science Foundation under Grant No. NSF PHY-1125915.

\appendix

\section{$U(N)$ duality.}
\label{sec:UN}

A generalization\cite{aharony2,seiberg2} of the dual description
(\ref{Lphi}) of a free Dirac fermion (\ref{Lpsi}) is the
$U(N)$-WFCS theory: \beq {\cal L} = |(\d_{\mu} - i a_{\mu})\phi| +
r |\phi|^2 + u |\phi|^4 + {\rm CS}_{U(N)}[a-A]_{1} + 2 N \cdot
{\rm CS}[g] \label{UN}\eeq where $\phi$ is an $N$-component
complex vector, $a$ is a $U(N)$ gauge field, $A$ as before is a
background $U(1)$ gauge field, and \beq {\rm CS}_{U(N)}[a]_1 =
\frac{i}{4\pi} {\rm tr}_{N}\left(a\wedge da - \frac{2 i}{3} a
\wedge a \wedge a\right) \label{CSU}\eeq The trace above is taken
in the fundamental representation of $U(N)$ and $a - A$ in
Eq.~(\ref{UN}) should be understood as $a - A\cdot \mathbbold{1}$,
where $\mathbbold{1}$ is the identity matrix. Surprisingly, the
theory (\ref{UN}) is conjectured to be dual to a single free Dirac
cone (\ref{Lpsi}) for {\it any} $N$.\cite{seiberg2} Again, the
symmetries and operators match: the electric current
$\bar{\psi}\gamma^{\mu} \psi$ maps to the flux current
$\frac{1}{4\pi} \epsilon^{\mu \nu \lambda} {\rm tr}(f_{\mu \nu})$.
Recalling that $\pi_1(U(N)) = Z$, we see that the quantization of
electric charge matches, and the fermion $\psi$ maps to the
monopole of $U(N)$. The phase diagram also matches. When $\phi$
is gapped, we have a $U(N)$ CS gauge theory at level $1$, which
carries no intrinsic topological order.  Integrating over $a$, we
find that $\sigma^A_{xy} = 0$. Moreover, since the chiral central
charge of $U(N)$ CS gauge theory at level $1$ is $c_- = -N$,  the
background gravitational CS term in Eq.~(\ref{UN}) is exactly
cancelled to give $c_- = 0$. On the other hand, when $r < 0$,
$\phi$ condenses and $U(N)$ is broken down to a $U(N-1)$ subgroup.
At low energies, we may keep only the components of $a$ in this
subgroup, so that $a$ is now an $N-1$ component Hermitian matrix.
Then ${\cal L}$ effectively takes the form: \beq {\cal L} = {\rm
CS}_{U(N-1)}[a-A]_{1} + \frac{i}{4\pi} A dA + 2N \cdot{\rm
CS}[g]\eeq Again, because the level of $a$ is $1$ , there is no
topological order. Integrating over $a$, we are left with
$\sigma^A_{xy} = 1$, and $c_- = N - (N-1) = 1$, $i.e.$ this phase is
a Chern insulator.

We can again obtain the theory (\ref{UN}) through a parton
construction. Represent the physical fermions $c_j = \sum_{\alpha}
b^{\dagger}_{\alpha,j} f_{\alpha,j}$ where $\alpha = 1\ldots N$.
There is a $U(N)$ gauge redundancy in this description: \beq
U(N)_a:\quad  b_j \rightarrow U_j b_j, \ \ \ f_j \rightarrow U_j
f_j\eeq which will lead to the emergence of a $U(N)$ gauge field
$a$. As before, we make $f$ carry the $U(1)_A$ global charge
(Eq.~(\ref{U1transf}), first line). Now, choose an identical
Chern-number $1$ band structure for each $f_{\alpha}$. Integrating
$f_\alpha$ out, we obtain the CS term for $a - A$ in
Eq.~(\ref{UN}) as well as a gravitational CS term with coefficient
$2 N$, corresponding to the chiral central charge $N$. The
slave-boson field $b$ can be in a gapped or condensed phase, the
transition between these phases is described by (\ref{UN}). While
we already discussed the nature of the two phases above, further
insight is provided by the edge structure. At the mean field
level, the edge has $N$ chiral $c = 1$ fermions. However, when $b$
is gapped, there are no $U(N)$ invariant degrees of freedom on the
edge, so after including fluctuations of $a$, the edge becomes a
trivial theory with $c  =0$. On the other hand, when $b$
condenses, for instance along the $N$th direction $\langle
b_\alpha \rangle \sim \delta_{\alpha, N}$, then the $U(N)$ gauge
group is broken down to $U(N-1)$. As a result, the $c  =1$ chiral
mode of $f_N$ now becomes a physical electron mode, while the
modes $f_{\alpha}$, $\alpha = 1\ldots N-1$ are still gapped by
fluctuations of $a$. Thus, the edge is now a $c  =1$ $A$-charged
mode, as one expects in a Chern insulator.

\section{$SO(N)$ Chern-Simons gauge theory.}
\label{sec:CS}
 In this appendix, we give a more careful definition
of the $SO(N)$ CS gauge theory in Eq.~(\ref{CSO}).\cite{DijkgraafW} Strictly
speaking, Eq.~(\ref{CSO}) is only meaningful when $a$ is a 1-form.
However, there can be non-trivial $SO(N)$ bundles over our
space-time manifold: we have to sum over such bundles. In general,
following our parton construction below Eq.~(\ref{parton1}), we
will define the CS action for the $SO(N)$ connection $a$ as the
partition function of $N$ identical copies of a $p+ip$
superconductor coupled to an $SO(N)$ gauge field. Representing the
$p+ip$ superconductor by a Majorana fermion $\chi$ with $m < 0$,
\beq \mathcal{L} = \bar{\chi}  (D_a + m) \chi \eeq where $\chi$ is
an $N$-color Majorana fermion, $D_a = \gamma^{\mu} (\d_{\mu} + i
\omega_{\mu} - i a_\mu)$, and $\omega_{\mu}$ is the spin
connection, we define:
\beqn \exp\left(-{\rm CS}_{SO(N)}[a]_1 - N \cdot {\rm CS}_g\right) &\equiv& \lim_{m \to \infty}\frac{Z_f(-m)}{Z_f(m)} \nn\\
&=& e^{-\pi i \eta(i D_a)} \label{etadef}\eeqn where \beq \eta =
\frac12 (\eta(0) + N_0) \label{eta0}\eeq and \beq \eta(s) =
\sum_{\lambda \neq 0} \rm{sgn}(\lambda) |\lambda|^{-s}\eeq where
$\lambda$'s are eigenvalues of $i D_a$, $N_0$ is the number of
zero modes of $i D_a$ and $\eta(0)$ is obtained by analytic
continuation from large real $s$. The ratio of partition functions
with $m < 0$ and $m > 0$ in (\ref{etadef}) is taken to cancel out
any ``non-topological" dependence of the action on the gauge field
and the metric.

Implicitly, in defining the Dirac operator $D_a$, we have given
the fermions $\chi$ not only an $SO(N)$ connection, but also a
spin-structure. In our parton construction (\ref{parton1}) such a
spin-structure will be inherited from the physical electron $c$.
In general, the action (\ref{etadef}) depends on the
spin-structure, which is a sign that the underlying microscopic
theory has a local electron operator.

We can use  Atiyah-Patodi-Singer (APS) theorem to rewrite  $\eta(i
D_a)$ as, \beqn \eta &=& \int_{X_4} \bigg(\frac{1}{2\cdot (2
\pi)^2} {\rm tr}_{SO(N)} f\wedge f \nn \\ &+& \frac{N}{8 \cdot
24\pi^2} {\rm tr} R\wedge R\bigg)\quad ({\rm mod}\,\,
2)\label{AR}\eeqn Here, $X_4$ is a four-dimensional (Euclidean)
manifold that extends our physical three-dimensional (Euclidean)
manifold $M$, $i.e.$ $\d X_4 = M$. Furthermore, both $a$ and the
spin-structure on $M$ are assumed to extend to $X_4$. $R$ is the
Riemann tensor on $X_4$. It follows from the APS theorem that the
right-hand-side of Eq.~(\ref{AR}) is independent of the particular
extension, $i.e.$ it vanishes mod $2$ when $X_4$ has no boundary.
Another way to see this without directly appealing to the APS
theorem is as follows. When $X_4$ is closed, we have \beq p_1 =
\frac{1}{2\cdot (2 \pi)^2} \int_{X_4} {\rm tr}_{SO(N)} f\wedge
f\eeq and \beq \sigma = -\frac{1}{ 24\pi^2} \int_{X_4}  {\rm tr}
R\wedge R \eeq $p_1$ is the Pontryagin number of the $SO(N)$
bundle over $X_4$. On a general manifold, it is an integer.
However, on a spin manifold it is an even integer. $\sigma$ is the
signature of the manifold. On a general manifold it is an integer.
However, on a spin manifold it is a multiple of $16$. This implies
that if our three-manifold $M$ is endowed with a spin-structure,
we can separately define \beq {\rm CS}_{SO(N)}[a]_1 = \frac{i}{2
(4\pi)}  \int_{X_4} {\rm tr}_{SO(N)}  f\wedge f  \label{CSObulk}
\eeq and \beq {\rm CS}_g = \frac{i}{192 \pi} \int_{X_4} {\rm tr} R
\wedge R \label{CSg} \eeq Neither (\ref{CSObulk}) nor (\ref{CSg})
depend on the extension to $X_4$ as long as the spin-structure is
also extended.  Eq.~(\ref{CSObulk}) is the standard definition of
$SO(N)_1$ gauge theory; for level $k$ - simply multiply
Eq.~(\ref{CSObulk}) by $k$. From the preceding discussion we see
that the action (\ref{CSObulk}) generally depends on the spin-structure for odd $k$, but not even $k$. Physically, when the
theory has bosonic $a$-matter fields, the odd $k$ theories always
have transparent fermions in the spectrum, while even $k$ theories
are consistent theories of microscopic bosons. In this paper, we
are principally interested in the case $k  = 1$: it is consistent
that our dual theory has a local fermion in its spectrum.

We note that for even $N$ and odd $k$ one can still make the
theory live in a purely bosonic Hilbert space, if one combines
both the CS term for $a$ and the gravitational CS term via the
definition (\ref{etadef}) or equivalently through Eq.~(\ref{AR}).
Physically, in this case one must restrict to fermionic $a$-matter
only. Indeed, for even $N$, the center of $SO(N)$ is $Z_2 = \{1,
-1\}$. The transition functions for the fermionic matter field
$\chi$ live in the group $(SO(N)\times Spin(3))/Z_2$, $i.e.$ $\chi$
feels only the combination of spin-structure and $SO(N)$
connection. Thus, if one includes only fermionic $a$-matter, one
does not need to give a spin-structure as an input and we get a
theory of microscopic bosons. However, if we allow for bosonic
matter $\phi$ (as our dual theory does), than the $\phi$ action
depends on the $SO(N)$ connection, but not on spin-structure, so
the entire action depends on both the $SO(N)$ connection and the
spin-structure.  So in this case, we still have microscopic
fermions in the Hilbert space.

We conclude by reviewing our notation for the case $N = 3$. In
this case, one may attempt to lift the $SO(3)$ bundle over $M$ to
an $SU(2)$ bundle. The lift does not always exist, so the two
theories are generally different in their global properties
(although local properties will be identical). When the lift does
exist, the $SO(3)$ action becomes, \beq {\rm CS}_{SO(3)}[a]_k = 2
\times \frac{i k}{4\pi} \int {\rm tr}_{SU(2)} \left(\hat{a} \wedge
d \hat{a}  - \frac{2 i}{3} \hat{a}\wedge \hat{a} \wedge
\hat{a}\right)\eeq Here the trace is over the spin $1/2$
representation of $SU(2)$ and $\hat{a}$ is the corresponding lift
of $a$. (We can write the action directly in 3d, because $SU(2)$
in $d = 3$ does not admit non-trivial bundles). That is $SO(3)_k$
goes to $SU(2)_{2k}$.

\section{$Z_2$ symmetry.}
\label{app:Z2}
In this section, we discuss the issue of the
``extra" global $Z_2$ symmetry of the dual theory (\ref{SON}) in
more detail. We will argue that this symmetry is realized in the
same way in both phases of (\ref{SON}), thus, it is natural that
the $Z_2$ gap remains finite across the transition.

It is useful to have an explicitly $Z_2$ symmetric parton
construction as a starting point for the analysis. Imagine we use
the parton decompostion \beq c = \sum_{\alpha = 1}^{N} \phi_\alpha
\chi_\alpha \label{partapp}\eeq with $c$ - the electron, $\phi_\alpha$ - slave
bosons and $\chi_\alpha$ - slave fermions. This decomposition has
an $O(N)$ gauge symmetry. As before, we place $\chi_\alpha$ into
identical $p_x + i p_y$ superconductor band structures, and
$\phi_\alpha$ into a trivial paramagnet. This yields the effective
theory (\ref{SON}) with an $O(N)$ gauge group. Now, imagine
further that the system has some Ising spin degrees of freedom
$\sigma$ (not related to the electrons $c$) that transform under a
global $Z_2$ symmetry: \beq Z_2:\,\,\, \sigma \to -\sigma, \quad c
\to c\eeq Imagine we condense the bosonic operator $\Phi \sigma$,
with  $\Phi$ defined by Eq.~(\ref{Phi}). This breaks the $O(N)$
gauge symmetry down to $SO(N)$. However, the physical global $Z_2$
symmetry is not broken, since its combination with an $O(N)$
reflection leaves $\Phi \sigma$ invariant. Thus, we conclude that
after $\Phi \sigma$ condensation, the global $Z_2$  symmetry acts
as an $O(N)$ reflection of $\phi_\alpha$ and $\chi_\alpha$. This
is precisely the ``extra" global symmetry of theory (\ref{SON}).

We would like to show that from the point of view of this global
$Z_2$ symmetry, both  $r > 0$ and $r < 0$ phases of (\ref{SON})
realize the same SPT. To argue this, let us look at the edge of
the system: a $Z_2$ SPT would possess non-trivial $Z_2$ carrying
gapless edge modes.\cite{levinguz8} At the mean field level,
$\chi_\alpha$ has $N$ chiral $c = 1/2$ modes $f_\alpha$
transforming under the $O(N)$ symmetry. However, after coupling to
the $SO(N)$ gauge field, all these modes are eliminated.
Therefore, the $r > 0$ phase must realize a trivial $Z_2$ SPT.
Let's now turn to $r < 0$. Suppose $\phi$ condenses along the
$N$'th direction, $\langle \phi_\alpha \rangle = \delta_{\alpha
N}$. The $SO(N)$ group is broken to $SO(N-1)$ acting on $\alpha =
1 \ldots N-1$. Moreover, the global $Z_2$ symmetry (after
potentially combining with an $SO(N)$ gauge rotation) now acts as
$O(N-1)$ on $\alpha = 1 \ldots N -1$. Therefore, the edge mode
$f_N$ is not coupled to the $SO(N-1)$ gauge field and, moreover,
is neutral under the global $Z_2$. On the other hand, the edge
modes $f_\alpha$ with $\alpha = 1\ldots N-1$ are eliminated by
fluctuations of the $SO(N-1)$ gauge field. So for $r < 0$, there
are no $Z_2$ carrying edge modes, which means that this phase is
also trivial from the $Z_2$ SPT standpoint.

Another way to diagnose the presence of a $Z_2$ SPT is to look at
the bulk response to $Z_2$ fluxes, $e.g.$ at the partition function
of the theory on some closed manifold in the presence of
background $Z_2$ gauge field $b$. Let us consider the case of odd
$N$, so $O(N) = Z_2 \times SO(N)$. For $r > 0$, the $Z_2$ symmetry
simply acts on partons $\chi_\alpha$ as $Z_2:\,\, \chi_\alpha \to
- \chi_\alpha$. This coincides with the action of fermion parity
on $\chi_\alpha$. But we know that our system is a trivial $c_-
=0$ superconductor. Therefore, its partition function does not
depend on spin-structure, and consequently, on the background
$Z_2$ gauge field $b$. So the $r > 0$ phase realizes a trivial
$Z_2$ SPT.

Next, we turn to the phase $r < 0$. As we already discussed, the
fermion $\chi_N$ now forms a $p_x+ip_y$ superconductor and is
neutral under $Z_2$. It, therefore, does not contribute any $b$
dependence to the partition function. We are thus left with the
contribution of fermions $\chi_\alpha$, $\alpha = 1\ldots N-1$
coupled to an $SO(N-1)$ gauge field and charged under the
$O(N-1)/SO(N-1)$ global symmetry. We note that because $N-1$ is
even (and all bosonic charge matter is trivial and gapped), this
system can effectively be thought as living in a bosonic
microscopic Hilbert space (see appendix~\ref{sec:CS}). Therefore,
as a $Z_2$ SPT, it must realize either a trivial phase or the
Chen-Liu-Wen-Levin-Gu (CLWLG) bosonic SPT phase.\cite{CZX,levingu} These can be distinguished by their
partition functions on $RP^3$ in the presence of a background
$Z_2$ flux. In section \ref{sec:O2}, we will explicitly compute
this partition function for $N -1 =2$ and show that, indeed, a
trivial $Z_2$ SPT phase is realized. For other $N$, we don't have
an explicit bulk partition function computation and have to rely
on the edge argument above.

Despite the above arguments, from a mean field viewpoint it
appears surprising that the operator (\ref{Phi}) can have
exponentially decaying correlation functions at the critical point
$r = 0$. We now give an example of a theory where a similar
phenomenon takes place. Consider the WFCS theory (\ref{Lphi}),
which is an $SO(2)$ cousin of our non-Abelian theory (\ref{SON}).
This theory has a $Z_2 = O(2)/SO(2)$ symmetry, which acts as $\phi
\to \phi^{\dagger}$. On the dual Dirac fermion side (\ref{Lpsi}),
this symmetry is precisely the charge-conjugation symmetry $Z_2:
\psi \to C \bar{\psi}^T$. The gap to this symmetry vanishes at the
critical point: for instance, the physical electric current
operator $\bar{\psi} \gamma^{\mu} \psi$ is odd under $C$. Now,
imagine perturbing the Dirac fermion by a superconducting mass
term (which breaks the global $U(1)$ symmetry):
\beqn H &=& \psi^{\dagger} (-i \d_x \sigma^x - i \d_y \sigma^z) \psi + m \psi^{\dagger} \sigma^y \psi +\nn\\
&+& \frac12 \Delta \psi^{\dagger} \sigma^y \psi^* + \frac12
\Delta^* \psi^T \sigma^y \psi \nn\eeqn Here, we've written the
real-time Hamiltonian and made a choice of $\gamma$ matrices. For
$\Delta = \Delta^*$, the charge-conjugation symmetry $Z_2: \psi
\to \psi^{\dagger}$ is preserved.
%On the WFCS side, the superconducting mass perturbation maps to a double instanton of gauge field $a$. In the Dirac theory,
We see that the Dirac transition splits into two Majorana
transitions. Indeed, writing $\psi = \xi_1 + i \xi_2$,
\beqn H &=& \xi^T_1 (-i \d_x \sigma^x - i \d_y \sigma^z) \xi_1
+ (m + \Delta) \xi^T_1 \sigma^y \xi_1\nn\\
&+& \xi^T_2 (-i \d_x \sigma^x - i \d_y \sigma^z) \xi_2 + (m -
\Delta) \xi^T_2 \sigma^y \xi_2 \nn\eeqn The Majorana fermion
$\xi_1$ becomes massless at $m = -\Delta$, and $\xi_2$ becomes
massless at $m = \Delta$. The $Z_2$ symmetry acts as, $\xi_1 \to
\xi_1, \,\, \xi_2 \to -\xi_2$. Since at finite $\Delta$, $\xi_1$
and $\xi_2$ are never simultaneously gapless, all bosons charged
under $Z_2$ remain gapped throughout the phase diagram. Now, on
the WFCS side, the superconducting mass perturbation maps to a
double monopole of $a$. Therefore, in this perturbed WFCS theory,
the $Z_2$ charge gap (for bosons) also never closes, just as we
expect in the non-Abelian theory (\ref{SON}). Of course, in the
perturbed Abelian theory,  an even more dramatic effect is that
the transition at $r = 0$ splits into two; in the non-Abelian case
(\ref{SON}) there is no reason to expect such a splitting. We note
that the perturbed Abelian WFCS theory can be thought as arising
by starting with the $N = 3$ non-Abelian WFCS theory (\ref{SON}),
going to the $r < 0$ phase where $SO(3)$ gauge symmetry is broken
to $SO(2)$; the Abelian WFCS transition then corresponds to
further breaking the $SO(2)$ gauge group to trivial. Note that in
such a construction, the $SO(2)$ theory will be naturally
perturbed by double monopole operators, as they are allowed in
the original $SO(3)$ theory. Our assumption is that the gap to
$Z_2 = O(3)/SO(3)$ bosons remains finite throughout the above
sequence of phase transitions.

\subsection{$O(2)_1$ theory.}
\label{sec:O2} In this section, we focus on the theory (\ref{SON})
with $N = 3$ in the regime $r < 0$ and compute its partition
function on $RP^3$ in the presence of a background $Z_2$ flux $b$.
This will serve as a check that this phase is not a $Z_2$ SPT.
Indeed, as we argued above, the $r < 0$ phase
(apart from a $Z_2$ neutral $p_x+ip_y$ superconductor) is at most
a CLWLG bosonic $Z_2$ SPT. The partition function of the
CLWLG phase on a manifold $M$ in the presence of a background
$Z_2$ flux $b \in H^1(M, Z_2)$ is given by\cite{DijkgraafW,kapustin2} \beq Z[b] =
\exp\left(\pi i \int_M b^3\right)\eeq On $RP^3$, $H^1(RP^3, Z_2) =
Z_2$, and for the non-trivial $b$, we have $Z[b] = -1$. On the
other hand, for a trivial $Z_2$ SPT, we will have $Z[b]=1$ on any
manifold.

From the parton construction below Eq.~(\ref{partapp}) the
background flux $b$ and the dynamical $SO(2)$ gauge field $a$
combine to an $O(2) = SO(2) \rtimes Z_2$ gauge field: the action
is the partition function of the partons $\chi_1$, $\chi_2$
coupled to this $O(2)$ gauge field - it is given by
Eq.~(\ref{etadef}). (Here, we ignore the third parton $\chi_3$,
which sees no dynamical gauge field or background $Z_2$ gauge
field). Thus, the partition function takes the form:

%In this section, we perform some extra explicit calculations that confirm that $O(2)_1$ theory is the same as a Toric code (+ electron).
%We define the partition function of the $O(2)_1$ theory by considering two copies of a $p+ip$ superconductor coupled to an $O(2)$ gauge field. For $O(2)_k$ theory we can just take $k$ copies of the above system (coupled to the same $O(2)$ gauge field). Represent a $p+ip$ superconductor by a massive Majorana,
%\beq L = \chi^T C^* \gamma^{\mu} (D_a + m) \chi\eeq
%where $D_a = \gamma^{\mu} (\d_{\mu} + i \omega_{\mu} + i a_\mu \tau^2)$ and $\tau^2$ acts on the two $O(2)$ colours. Integrating the Majorana's out %and taking the ratio of partition function at $m > 0$  and $m < 0$, we obtain
%\beq Z[a] = \exp(\pi i k \eta[a])\eeq where
%\beq \eta = \frac12 (\eta(0) + N_0)\eeq
%and
%\beq \eta(s) = \sum_{\lambda_{\neq 0}} \rm{sgn}(\lambda) |\lambda|^{-s}\eeq
%with $\lambda$'s - eigenvalues of $i D_a$ (importantly, these eigenvalues are always doubly degenerate due to $[ i D_a, \sigma^2 K] = 0$). By usual arguments, small variations of $Z[a]$ are just given by usual CS term,
%\beq \frac{Z[a + \delta a]}{Z[a]} = \exp\left(\frac{i k}{4\pi} \int \delta a \wedge d a\right)\eeq

%The partition function of the $O(2)_k$ theory is then the integral of $Z[a]$ over $O(2)$ gauge fields $a$. We can break up this integral in the following way:
\beq Z[b] = \frac{1}{Vol({\cal G})} \sum_{H^2(M, \tilde{Z}_b)}
\int D a\, e^{-\pi i  \eta[a]}\eeq Once the $Z_2$ gauge field $b$
is fixed, $a$ becomes a $u(1)$ gauge field, but its transition
functions satisfy a cocycle condition twisted by $b$.
Topologically distinct $u(1)$ gauge fields correspond to elements
of $H^2(X, \tilde{Z}_b)$, where the coefficients are in the local
system twisted by $b$. The path integral is over gauge fields $a$
in each such class and the factor in front, $Vol({\cal G})^{-1}$,
corresponds to the volume of (twisted) $u(1)$ gauge group.
$\eta[a]$ is the $\eta$-invariant (\ref{eta0}) of the Dirac
operator in the background of the $O(2)$ gauge field $a$. A
standard computation gives,

\beq Z[b] =  \frac{ e^{2\pi i \eta(*d)/8} }{\sqrt{|T|}}
\sum_{a_{cl} \in T} e^{-\pi i  \eta[a_{cl}]} \label{eq:Zfinal}\eeq
Here $a_{cl}$ are topologically non-trivial flat $u(1)$ bundles
twisted by $b$ - these are in one-to-one correspondence with the
torsion subgroup $T = Tor(H^2(M, \tilde{Z}_b))$ and $|T|$ denotes
the order of $T$.  $\eta(*d)$ is the $\eta$-invariant of the
operator $* d$ acting on (twisted) 1-forms and 3-forms - it
vanishes for manifolds with an orientation-reversing isometry,
such as $RP^3$.

Let's compute the partition function on $RP^3$. As already noted,
$H^1(RP^3, Z_2) = Z_2$, so there is only one non-trivial $b$-flux
sector. In the untwisted ($b = 0$) sector, we have $H^2(RP^3, Z) =
Z_2$ which correspond to two spin-structures for slave-fermions
$\chi_\alpha$ on $RP^3$. Explicitly, working on the double-cover
$S^3$ and using the stereographic coordinates,
$(\frac{4\vec{u}}{u^2+4}, \frac{u^2 - 4}{u^2+4})$, we have under
the anti-podal map: \beq \chi_\alpha(\vec{u}) = \pm \frac{i u^i
\gamma^i}{|\vec{u}|}
\chi_\alpha\left(\frac{-4\vec{u}}{u^2}\right)\eeq Note that
$\chi_1$ and $\chi_2$ have the same spin-structure. Now, one of
the spin-structures has $\eta = 1/8$ and the other $\eta = -1/8$
(for a single fermion, say $\chi_1$). Adding contributions of
$\chi_1$ and $\chi_2$,  $\eta = \pm 1/4$. So, we get \beq
Z_{0}(RP^3) = \frac{1}{\sqrt{2}} (e^{\pi i/4} + e^{-\pi i/4}) =
1\eeq This is the correct partition function of a trivial
superconductor on $RP^3$, as needed.

Now, in the twisted sector, $H^2(RP^3, \tilde{Z}) = Z_1$, there is
just a single twisted $u(1)$ bundle that can be chosen to be: \beq
\chi_\alpha (\vec{u}) = \tau^z_{\alpha \beta} \frac{i u^i
\gamma^i}{|\vec{u}|} \chi_\beta
\left(\frac{-4\vec{u}}{u^2}\right)\eeq So $\chi_1$ and $\chi_2$
now have opposite spin-structures, so together they give $\eta =
0$. So, we get \beq Z_{{\rm twist}}(RP^3) = 1\eeq Since the
partition function in the twisted sector is trivial, this phase
must realize a trivial $Z_2$ SPT as claimed.

%\end{appendix}

\bibliography{majboson}

\begin{thebibliography}{33}
\expandafter\ifx\csname natexlab\endcsname\relax\def\natexlab#1{#1}\fi
\expandafter\ifx\csname bibnamefont\endcsname\relax
  \def\bibnamefont#1{#1}\fi
\expandafter\ifx\csname bibfnamefont\endcsname\relax
  \def\bibfnamefont#1{#1}\fi
\expandafter\ifx\csname citenamefont\endcsname\relax
  \def\citenamefont#1{#1}\fi
\expandafter\ifx\csname url\endcsname\relax
  \def\url#1{\texttt{#1}}\fi
\expandafter\ifx\csname urlprefix\endcsname\relax\def\urlprefix{URL }\fi
\providecommand{\bibinfo}[2]{#2}
\providecommand{\eprint}[2][]{\url{#2}}

\bibitem[{\citenamefont{Son}(2015)}]{son2015}
\bibinfo{author}{\bibfnamefont{D.~T.} \bibnamefont{Son}},
  \bibinfo{journal}{Phys. Rev. X} \textbf{\bibinfo{volume}{5}},
  \bibinfo{pages}{031027} (\bibinfo{year}{2015}).

\bibitem[{\citenamefont{Wang and Senthil}(2015)}]{wangsenthil1}
\bibinfo{author}{\bibfnamefont{C.}~\bibnamefont{Wang}} \bibnamefont{and}
  \bibinfo{author}{\bibfnamefont{T.}~\bibnamefont{Senthil}},
  \bibinfo{journal}{Phys. Rev. X} \textbf{\bibinfo{volume}{5}},
  \bibinfo{pages}{041031} (\bibinfo{year}{2015}).

\bibitem[{\citenamefont{Metlitski and Vishwanath}(2016)}]{maxashvin}
\bibinfo{author}{\bibfnamefont{M.~A.} \bibnamefont{Metlitski}}
  \bibnamefont{and}
  \bibinfo{author}{\bibfnamefont{A.}~\bibnamefont{Vishwanath}},
  \bibinfo{journal}{Phys. Rev. B} \textbf{\bibinfo{volume}{93}},
  \bibinfo{pages}{245151} (\bibinfo{year}{2016}).

\bibitem[{\citenamefont{Mross et~al.}(2016)\citenamefont{Mross, Alicea, and
  Motrunich}}]{mross}
\bibinfo{author}{\bibfnamefont{D.~F.} \bibnamefont{Mross}},
  \bibinfo{author}{\bibfnamefont{J.}~\bibnamefont{Alicea}}, \bibnamefont{and}
  \bibinfo{author}{\bibfnamefont{O.~I.} \bibnamefont{Motrunich}},
  \bibinfo{journal}{Phys. Rev. Lett.} \textbf{\bibinfo{volume}{117}},
  \bibinfo{pages}{016802} (\bibinfo{year}{2016}).

\bibitem[{\citenamefont{Karch and Tong}(2016)}]{karchtong}
\bibinfo{author}{\bibfnamefont{A.}~\bibnamefont{Karch}} \bibnamefont{and}
  \bibinfo{author}{\bibfnamefont{D.}~\bibnamefont{Tong}},
  \bibinfo{journal}{arXiv:1606.01893}  (\bibinfo{year}{2016}).

\bibitem[{\citenamefont{Seiberg et~al.}(2016)\citenamefont{Seiberg, Senthil,
  Wang, and Witten}}]{seiberg1}
\bibinfo{author}{\bibfnamefont{N.}~\bibnamefont{Seiberg}},
  \bibinfo{author}{\bibfnamefont{T.}~\bibnamefont{Senthil}},
  \bibinfo{author}{\bibfnamefont{C.}~\bibnamefont{Wang}}, \bibnamefont{and}
  \bibinfo{author}{\bibfnamefont{E.}~\bibnamefont{Witten}},
  \bibinfo{journal}{arXiv:1606.01989}  (\bibinfo{year}{2016}).

\bibitem[{\citenamefont{Peskin}(1978)}]{peskindual}
\bibinfo{author}{\bibfnamefont{M.~E.} \bibnamefont{Peskin}},
  \bibinfo{journal}{Annals Phys.} \textbf{\bibinfo{volume}{122}},
  \bibinfo{pages}{113} (\bibinfo{year}{1978}).

\bibitem[{\citenamefont{Dasgupta and Halperin}(1981)}]{halperindual}
\bibinfo{author}{\bibfnamefont{C.}~\bibnamefont{Dasgupta}} \bibnamefont{and}
  \bibinfo{author}{\bibfnamefont{B.~I.} \bibnamefont{Halperin}},
  \bibinfo{journal}{Phys. Rev. Lett.} \textbf{\bibinfo{volume}{47}},
  \bibinfo{pages}{1556} (\bibinfo{year}{1981}).

\bibitem[{\citenamefont{Polyakov}(1988)}]{PolyakovAM}
\bibinfo{author}{\bibfnamefont{A.~M.} \bibnamefont{Polyakov}},
  \bibinfo{journal}{Mod. Phys. Lett. A} \textbf{\bibinfo{volume}{3}},
  \bibinfo{pages}{325} (\bibinfo{year}{1988}).

\bibitem[{\citenamefont{Chen et~al.}(1993)\citenamefont{Chen, Fisher, and
  Wu}}]{wufisher}
\bibinfo{author}{\bibfnamefont{W.}~\bibnamefont{Chen}},
  \bibinfo{author}{\bibfnamefont{M.~P.~A.} \bibnamefont{Fisher}},
  \bibnamefont{and} \bibinfo{author}{\bibfnamefont{Y.-S.} \bibnamefont{Wu}},
  \bibinfo{journal}{Phys. Rev. B} \textbf{\bibinfo{volume}{48}},
  \bibinfo{pages}{13749} (\bibinfo{year}{1993}).

\bibitem[{\citenamefont{Barkeshli and McGreevy}(2014)}]{barkeshlimcgreevy}
\bibinfo{author}{\bibfnamefont{M.}~\bibnamefont{Barkeshli}} \bibnamefont{and}
  \bibinfo{author}{\bibfnamefont{J.}~\bibnamefont{McGreevy}},
  \bibinfo{journal}{Phys. Rev. B} \textbf{\bibinfo{volume}{89}},
  \bibinfo{pages}{235116} (\bibinfo{year}{2014}).

\bibitem[{\citenamefont{Giombi et~al.}(2012)\citenamefont{Giombi, Minwalla,
  Prakash, Trivedi, and Wadia}}]{minwalla}
\bibinfo{author}{\bibfnamefont{S.}~\bibnamefont{Giombi}},
  \bibinfo{author}{\bibfnamefont{S.}~\bibnamefont{Minwalla}},
  \bibinfo{author}{\bibfnamefont{S.}~\bibnamefont{Prakash}},
  \bibinfo{author}{\bibfnamefont{S.~P.} \bibnamefont{Trivedi}},
  \bibnamefont{and} \bibinfo{author}{\bibfnamefont{S.~R.} \bibnamefont{Wadia}},
  \bibinfo{journal}{Eur.Phys.J.} \textbf{\bibinfo{volume}{C72}},
  \bibinfo{pages}{2112} (\bibinfo{year}{2012}).

\bibitem[{\citenamefont{Aharony
  et~al.}(2012{\natexlab{a}})\citenamefont{Aharony, Gur-Ari, and
  Yacoby}}]{aharony1}
\bibinfo{author}{\bibfnamefont{O.}~\bibnamefont{Aharony}},
  \bibinfo{author}{\bibfnamefont{G.}~\bibnamefont{Gur-Ari}}, \bibnamefont{and}
  \bibinfo{author}{\bibfnamefont{R.}~\bibnamefont{Yacoby}},
  \bibinfo{journal}{JHEP} \textbf{\bibinfo{volume}{1203}}, \bibinfo{pages}{037}
  (\bibinfo{year}{2012}{\natexlab{a}}).

\bibitem[{\citenamefont{Aharony
  et~al.}(2012{\natexlab{b}})\citenamefont{Aharony, Gur-Ari, and
  Yacoby}}]{aharony2}
\bibinfo{author}{\bibfnamefont{O.}~\bibnamefont{Aharony}},
  \bibinfo{author}{\bibfnamefont{G.}~\bibnamefont{Gur-Ari}}, \bibnamefont{and}
  \bibinfo{author}{\bibfnamefont{R.}~\bibnamefont{Yacoby}},
  \bibinfo{journal}{JHEP} \textbf{\bibinfo{volume}{1212}}, \bibinfo{pages}{028}
  (\bibinfo{year}{2012}{\natexlab{b}}).

\bibitem[{\citenamefont{Aharony}(2016)}]{aharony3}
\bibinfo{author}{\bibfnamefont{O.}~\bibnamefont{Aharony}},
  \bibinfo{journal}{JHEP} \textbf{\bibinfo{volume}{1602}}, \bibinfo{pages}{093}
  (\bibinfo{year}{2016}).

\bibitem[{\citenamefont{Hsin and Seiberg}(2016)}]{seiberg2}
\bibinfo{author}{\bibfnamefont{P.-S.} \bibnamefont{Hsin}} \bibnamefont{and}
  \bibinfo{author}{\bibfnamefont{N.}~\bibnamefont{Seiberg}},
  \bibinfo{journal}{arXiv:1607.07457}  (\bibinfo{year}{2016}).

\bibitem[{\citenamefont{Jenquin}(2005)}]{Jenquin1}
\bibinfo{author}{\bibfnamefont{J.~A.} \bibnamefont{Jenquin}}
  (\bibinfo{year}{2005}), \eprint{arXiv:math/0504524}.

\bibitem[{\citenamefont{Jenquin}(2006)}]{Jenquin2}
\bibinfo{author}{\bibfnamefont{J.~A.} \bibnamefont{Jenquin}}
  (\bibinfo{year}{2006}), \eprint{arXiv:math/0605239}.

\bibitem[{\citenamefont{Kapustin et~al.}(2015)\citenamefont{Kapustin,
  Thorngren, Turzillo, and Wang}}]{KapustinF}
\bibinfo{author}{\bibfnamefont{A.}~\bibnamefont{Kapustin}},
  \bibinfo{author}{\bibfnamefont{R.}~\bibnamefont{Thorngren}},
  \bibinfo{author}{\bibfnamefont{A.}~\bibnamefont{Turzillo}}, \bibnamefont{and}
  \bibinfo{author}{\bibfnamefont{Z.}~\bibnamefont{Wang}},
  \bibinfo{journal}{JHEP} \textbf{\bibinfo{volume}{1512}}, \bibinfo{pages}{052}
  (\bibinfo{year}{2015}).

\bibitem[{\citenamefont{Grover and Vishwanath}(2012)}]{grovervishwanath1}
\bibinfo{author}{\bibfnamefont{T.}~\bibnamefont{Grover}} \bibnamefont{and}
  \bibinfo{author}{\bibfnamefont{A.}~\bibnamefont{Vishwanath}},
  \bibinfo{journal}{arXiv:1206.1332}  (\bibinfo{year}{2012}).

\bibitem[{\citenamefont{Grover et~al.}(2014)\citenamefont{Grover, Sheng, and
  Vishwanath}}]{grovervishwanath2}
\bibinfo{author}{\bibfnamefont{T.}~\bibnamefont{Grover}},
  \bibinfo{author}{\bibfnamefont{D.~N.} \bibnamefont{Sheng}}, \bibnamefont{and}
  \bibinfo{author}{\bibfnamefont{A.}~\bibnamefont{Vishwanath}},
  \bibinfo{journal}{Science} \textbf{\bibinfo{volume}{344}},
  \bibinfo{pages}{280} (\bibinfo{year}{2014}).

\bibitem[{\citenamefont{Fei et~al.}(2016)\citenamefont{Fei, Giombi, Klebanov,
  and Tarnopolsky}}]{klebanov}
\bibinfo{author}{\bibfnamefont{L.}~\bibnamefont{Fei}},
  \bibinfo{author}{\bibfnamefont{S.}~\bibnamefont{Giombi}},
  \bibinfo{author}{\bibfnamefont{I.~R.} \bibnamefont{Klebanov}},
  \bibnamefont{and}
  \bibinfo{author}{\bibfnamefont{G.}~\bibnamefont{Tarnopolsky}},
  \bibinfo{journal}{arXiv:1607.05316}  (\bibinfo{year}{2016}).

\bibitem[{\citenamefont{Fidkowski et~al.}(2013)\citenamefont{Fidkowski, Chen,
  and Vishwanath}}]{fidkowskihe}
\bibinfo{author}{\bibfnamefont{L.}~\bibnamefont{Fidkowski}},
  \bibinfo{author}{\bibfnamefont{X.}~\bibnamefont{Chen}}, \bibnamefont{and}
  \bibinfo{author}{\bibfnamefont{A.}~\bibnamefont{Vishwanath}},
  \bibinfo{journal}{Phys. Rev. X} \textbf{\bibinfo{volume}{3}},
  \bibinfo{pages}{041016} (\bibinfo{year}{2013}).

\bibitem[{\citenamefont{Sahoo et~al.}(2016)\citenamefont{Sahoo, Zhang, and
  Teo}}]{Teo}
\bibinfo{author}{\bibfnamefont{S.}~\bibnamefont{Sahoo}},
  \bibinfo{author}{\bibfnamefont{Z.}~\bibnamefont{Zhang}}, \bibnamefont{and}
  \bibinfo{author}{\bibfnamefont{J.~C.~Y.} \bibnamefont{Teo}},
  \bibinfo{journal}{Phys. Rev. B} \textbf{\bibinfo{volume}{94}},
  \bibinfo{pages}{165142} (\bibinfo{year}{2016}).

\bibitem[{\citenamefont{Wang and Levin}(2016)}]{WangLevin}
\bibinfo{author}{\bibfnamefont{C.}~\bibnamefont{Wang}} \bibnamefont{and}
  \bibinfo{author}{\bibfnamefont{M.}~\bibnamefont{Levin}},
  \bibinfo{journal}{arXiv:1610.08478}  (\bibinfo{year}{2016}).

\bibitem[{\citenamefont{Chen et~al.}(2014)\citenamefont{Chen, Fidkowski, and
  Vishwanath}}]{TI_fidkowski2}
\bibinfo{author}{\bibfnamefont{X.}~\bibnamefont{Chen}},
  \bibinfo{author}{\bibfnamefont{L.}~\bibnamefont{Fidkowski}},
  \bibnamefont{and}
  \bibinfo{author}{\bibfnamefont{A.}~\bibnamefont{Vishwanath}},
  \bibinfo{journal}{Phys. Rev. B} \textbf{\bibinfo{volume}{89}},
  \bibinfo{pages}{165132} (\bibinfo{year}{2014}).

\bibitem[{\citenamefont{Bonderson et~al.}(2013)\citenamefont{Bonderson, Nayak,
  and Qi}}]{Ti_qi}
\bibinfo{author}{\bibfnamefont{P.}~\bibnamefont{Bonderson}},
  \bibinfo{author}{\bibfnamefont{C.}~\bibnamefont{Nayak}}, \bibnamefont{and}
  \bibinfo{author}{\bibfnamefont{X.-L.} \bibnamefont{Qi}}, \bibinfo{journal}{J.
  Stat. Mech.} p. \bibinfo{pages}{P09016} (\bibinfo{year}{2013}).

\bibitem[{\citenamefont{Dijkgraaf and Witten}(1990)}]{DijkgraafW}
\bibinfo{author}{\bibfnamefont{R.}~\bibnamefont{Dijkgraaf}} \bibnamefont{and}
  \bibinfo{author}{\bibfnamefont{E.}~\bibnamefont{Witten}},
  \bibinfo{journal}{Commun. Math. Phys.} \textbf{\bibinfo{volume}{129}},
  \bibinfo{pages}{393} (\bibinfo{year}{1990}).

\bibitem[{\citenamefont{Gu and Levin}(2013)}]{levinguz8}
\bibinfo{author}{\bibfnamefont{Z.-C.} \bibnamefont{Gu}} \bibnamefont{and}
  \bibinfo{author}{\bibfnamefont{M.}~\bibnamefont{Levin}},
  \bibinfo{journal}{arXiv:1304.4569}  (\bibinfo{year}{2013}).

\bibitem[{\citenamefont{Chen et~al.}(2011)\citenamefont{Chen, Liu, and
  Wen}}]{CZX}
\bibinfo{author}{\bibfnamefont{X.}~\bibnamefont{Chen}},
  \bibinfo{author}{\bibfnamefont{Z.-X.} \bibnamefont{Liu}}, \bibnamefont{and}
  \bibinfo{author}{\bibfnamefont{X.-G.} \bibnamefont{Wen}},
  \bibinfo{journal}{Phys. Rev. B} \textbf{\bibinfo{volume}{84}},
  \bibinfo{pages}{235141} (\bibinfo{year}{2011}).

\bibitem[{\citenamefont{Levin and Gu}(2012)}]{levingu}
\bibinfo{author}{\bibfnamefont{M.}~\bibnamefont{Levin}} \bibnamefont{and}
  \bibinfo{author}{\bibfnamefont{Z.-C.} \bibnamefont{Gu}},
  \bibinfo{journal}{Phys. Rev. B} \textbf{\bibinfo{volume}{86}},
  \bibinfo{pages}{115109} (\bibinfo{year}{2012}).

\bibitem[{\citenamefont{Kapustin and Thorngren}(2014)}]{kapustin2}
\bibinfo{author}{\bibfnamefont{A.}~\bibnamefont{Kapustin}} \bibnamefont{and}
  \bibinfo{author}{\bibfnamefont{R.}~\bibnamefont{Thorngren}},
  \bibinfo{journal}{arXiv:1404.3230}  (\bibinfo{year}{2014}).

\bibitem[{\citenamefont{Vicari}(2007)}]{vicari2007}
\bibinfo{author}{\bibfnamefont{E.}~\bibnamefont{Vicari}},
  \bibinfo{journal}{PoS(Lattie 2007):023}  (\bibinfo{year}{2007}).

\end{thebibliography}

\end{document}